\documentclass[%
aps,%
pre,%
reprint,%
showpacs,%
amsmath,%
amssymb,%
nofootinbib%
]{revtex4-1}

\usepackage{amsfonts}
\usepackage{graphicx}
\usepackage{xcolor}
\usepackage{mathtools}
\usepackage{bm}
\usepackage{dsfont}
\usepackage{hyperref}

\newcommand{\be}{\begin{equation}}
\newcommand{\ee}{\end{equation}}

\begin{document}

\title{Functional Renormalization Group approach to the Kraichnan model}

\author{Carlo Pagani}
\email[]{capagani@uni-mainz.de}
\affiliation{Institut f\"{u}r Physik (WA THEP) Johannes-Gutenberg-Universit\"{a}t, Staudingerweg 7, 55099 Mainz, Germany}


\begin{abstract}
We study the anomalous scaling of the structure functions of a scalar field advected by a random Gaussian velocity field, the Kraichnan model,
by means of Functional Renormalization Group techniques. We analyze the symmetries of the model and derive the leading correction to the structure functions considering the renormalization of composite operators and applying the operator product expansion.
\end{abstract}

\maketitle


\section{Introduction \label{sec:Introduction}}

Understanding the scaling of the velocity structure functions in turbulent
flows is a challenging problem. Such scaling is known to be universal
in the inertial range and shows small departures from the behaviour
predicted by Kolmogorov \cite{Kolmogorov_K41}. To make progress in understanding
the anomalous scaling of the structure functions a model for an advected
passive scalar field, the Kraichnan model \cite{Kraichnan:1994zz}, turned
out to be illuminating (similar models were considered long time ago
by Corssin and Obukhov \cite{Corssin,Obukhov}). One of the features which
makes this model appealing is that the corrections to the canonical
scaling of the structure functions can be analytically computed considering
the zero modes of some operator ${\cal M}_{n}^{*}$ which acts on
the equal time correlation functions of the scalar field \cite{Gawedzki:1995zz,Chertkov:1995wi}.
This result has been obtained via various methods among which 
the expansion in the H{\"{o}}lder exponent appearing in the covariance of the velocity field 
(i.e.: the $\varepsilon$ appearing in equation (\ref{eq:def_Dv_expl})), 
an expansion for large spatial dimension and the Batchelor limit ($\varepsilon$ close to $2$)
\cite{Gawedzki:1995zz,Chertkov:1995wi,Chertkov:1995zz,ShraimanSiggia:1995,Bernard:1996um}.
Furthermore the zero modes have been understood as statistical 
conservation laws \cite{BernardGawedzkiKupiainen}.
For a review of the results and a complete discussion
we refer to \cite{Falkovich:2001zz}. In this work we study the scaling of the structure functions of the
Kraichnan model by means
of the Functional Renormalization Group which offers the advantage of being a very adaptable framework.

The Renormalization Groups proved to be ideal to understand
the anomalous scaling in statistical systems but the applications
of its inherent concepts go far beyond. The Functional
Renormalization Group (FRG) is a functional realization of the Wilsonian renormalization program
and is applied successfully to many problems
regarding out of equilibrium systems, for a review see \cite{Berges:2012ty}
and \cite{Canet:2011wf}. In the FRG framework one considers a scale
dependent effective action, the Effective Average Action (EAA), which
interpolates between the classical action and the standard effective
action \cite{Wetterich:1992yh,Ellwanger:1993mw,Morris:1993qb}. Within this framework the issue of
a fixed point for the functional integral associated to the randomly
stirred Navier-Stokes equation has been first considered in \cite{Collina:1997jc}
and more recently in \cite{MonasterioGinanneschi2012} and \cite{Canet:2014cta,Canet:2014dta}. 
In particular in \cite{Canet:2014dta} a fixed point solution of the FRG equation associated
to this problem has been investigated in $d=2$ and $d=3$. Nevertheless
there is not still a quantitative prediction for the anomalous exponent
of the structure functions and it may turn useful to address this
question in a simpler setting. Let us stress that the FRG has been
proved to be a solid framework in similar problems such as the scaling
exponents of the Kardar-Parisi-Zhang equation and its correlation
functions in the strong coupling regime \cite{Canet:2011ez,Kloss:2013xva} 
(see \cite{Mathey:2014xxa} for the Burgers' equation). \\
$ \mbox{   }$
The RG analysis of the Kraichnan model has been performed in a perturbative
framework by means of an $\varepsilon-$expansion which is very similar
to the one used in the analysis of critical phenomena \cite{Adzhemyan1998_incompressible}.
The $\varepsilon-$expansion framework has been applied also to many
other problems ranging from the randomly stirred Navier-Stokes equation
to magnetic hydrodynamics, we refer to \cite{Adzhemyan1996_review} and \cite{AdzhemyanBook1999}
for a complete list of references. Actually this framework turned
out to be very efficient for the Kraichnan model and the anomalous exponents of the structure
function have been computed up to order $O\left(\varepsilon^{3}\right)$
\cite{Adzhemyan2001_eps3}. 
The inverse and the Wilsonian RG and their connection to the
zero modes results have been studied in \cite{Kupiainen:2006em}.
In this work we analyze the FRG approach to the Kraichnan
model; this allows us to investigate a number of issues in a simpler setting and shows that the FRG is a viable
framework for the computation of the anomalous exponents for the structure
functions. Thus the main purpose of this work is to study in a concrete example
the techniques required to understand the scaling of the structure
functions in the FRG framework and in particular the role of composite
operators. The paper is organized as follows: in the section \ref{sec:Path-integral-formulation}
we introduce the path integral formalism for the model. 
In section \ref{sec:Functional-Renormalization-Group}
we consider the relevant RG equations and show how to obtain the latter ones
within the FRG approach. In section \ref{sec:The-IR-fixed} we study
the IR fixed point of the model while in section \ref{sec:RG-flow-of_composite_operators}
we consider the anomalous dimension of the relevant composite operators.
In section \ref{sec:Scaling-of-the-Sn} we put together our results
and discuss the scaling of the structure functions. In the final section
we summarize our findings and consider possible outlooks.

\section{Path integral formulation of the Kraichnan model \label{sec:Path-integral-formulation}}

In this section we recall the definition of the Kraichnan model and
derive its path integral representation. 
The symmetries of the model are described in appendix \ref{sub:Symmetries} together with some non-renormalization
theorems which follow. 
Let us consider a scalar quantity $\theta\left(t,x\right)$,
e.g.: the temperature, which is advected by a fluid having random
velocity $v\left(t,x\right)$. 
The equation for the dynamics of $\theta\left(t,x\right)$ is
\begin{eqnarray}
\partial_{t}\theta+v^{i}\partial_{i}\theta-\frac{\kappa}{2}\partial^{2}\theta & = & f\label{eq:EOM_theta_force}
\end{eqnarray}
where $\kappa$ is the molecular diffusivity and $f$ is a stochastic
forcing term with Gaussian statistic and zero average. Note that equation (\ref{eq:EOM_theta_force}) is a stochastic
differential equation with multiplicative noise and is interpreted in the Stratonovich sense.
More precisely the covariance of the forcing term is:
\begin{eqnarray*}
\langle f\left(t_{1},x_{1}\right)f\left(t_{2},x_{2}\right)\rangle & = & \delta\left(t_{1}-t_{2}\right)F\left(\frac{x_{1}-x_{2}}{L_{F}}\right)\\
 & \equiv & D_{\bar{\theta}} 
\end{eqnarray*}
where $L_{F}$ is the characteristic scale of the forcing (eventually
we will denote $L_{F}^{-1}=M$). The random velocity $v\left(t,x\right)$
is also Gaussian and has zero average:
\begin{eqnarray}
\langle v_i\left(t_{1},x_{1}\right)v_j\left(t_{2},x_{2}\right)\rangle & = & \frac{D_{0}}{\left(2\pi\right)^{d}}\delta\left(t_1-t_2\right)\int d^{d}k e^{ikx_{1}-ikx_{2}} \nonumber \\ 
& \, & \,\,\, \frac{P_{ij}}{\left(k^{2}+m^{2}\right)^{\frac{d}{2}+\frac{\varepsilon}{2}}} \label{eq:def_Dv_expl} \\ 
 & \equiv & D_{v} \label{eq:def_Dv}
\end{eqnarray}
where
\begin{eqnarray*}
P_{ij} & = & \delta_{ij}-\frac{k_{i}k_{j}}{k^{2}}
\end{eqnarray*}
is the projector onto the transversal components (the fluid is incompressible).
Throughout this work we assume the Stratonovich discretization prescription for 
all the relavant formulas unless otherwise stated.
Note that the mass $m$ in the denominator avoids IR divergences and
is a further scale besides $L_{F}$, which appears in the force correlator.
Actually in the inertial regime the structure functions do not depend on $m$ 
and we set $m$ to zero (if one wanted to it is easy to keep this parameter). We will not need to introduce
$m$ at all since the FRG framework is safe from IR divergences by construction.

Averages can be computed via the following path integral:
\begin{eqnarray*}
Z & = & \int{\cal D}f{\cal D}v{\cal D}\theta P_{ff}P_{vv}\delta\left(\theta-\theta_{0}\right)
\end{eqnarray*}
where $\theta_{0}$ is a solution of equation (\ref{eq:EOM_theta_force})
and $P_{ff}$ and $P_{vv}$ are the Gaussian weighting factors of $f$
and $v$ respectively. We have:
\begin{eqnarray}
Z & = & \int{\cal D}f{\cal D}v{\cal D}\theta P_{ff}P_{vv}\delta\left(\partial_{t}\theta+v^{i}\partial_{i}\theta-\frac{\kappa}{2}\partial^{2}\theta-f\right) \nonumber \\
\, &\, & \, \left\Vert \det\left(\partial_{t}+v^{i}\partial_{i}-\frac{\kappa}{2}\partial^{2}\right)\right\Vert \label{eq:Z_and_det}
\end{eqnarray}
where the determinant comes from the change of variables in the path
integral. For the time being we shall neglect this determinant and we
will come back to it in a moment:
\begin{eqnarray*}
Z & = & \int{\cal D}f{\cal D}v{\cal D}\theta e^{-\frac{1}{2}\int f\langle ff\rangle^{-1} f}e^{-\frac{1}{2}\int v\langle vv\rangle^{-1}v} \\
&\, & \,\, \delta\left(\partial_{t}\theta+v^{i}\partial_{i}\theta-\frac{\kappa}{2}\partial^{2}\theta-f\right)\\
 & = & \int{\cal D}\bar{\theta}{\cal D}f{\cal D}v{\cal D}\theta e^{-\frac{1}{2}\int f\langle ff\rangle^{-1} f}e^{-\frac{1}{2}v\langle vv\rangle^{-1}v} \\
&\, & \,\, \exp\left[i\int\bar{\theta}\left(\partial_{t}\theta+v^{i}\partial_{i}\theta-\frac{\kappa}{2}\partial^{2}\theta-f\right)\right]\\
 & = & \int{\cal D}\bar{\theta}{\cal D}v{\cal D}\theta e^{+\frac{1}{2}\int\bar{\theta}\langle ff\rangle\bar{\theta}}e^{-\frac{1}{2}\int v\langle vv\rangle^{-1}v} \\
&\, & \,\, \exp\left[-\int\bar{\theta}\left(\partial_{t}\theta+v^{i}\partial_{i}\theta-\frac{\kappa}{2}\partial^{2}\theta\right)\right]\,.
\end{eqnarray*}
In the last line we redefined the field $\bar{\theta}$ via $\bar{\theta}\rightarrow i \bar{\theta}$.
The determinant in equation (\ref{eq:Z_and_det}) is very similar to the one encountered in the path
integral representation of the Langevin equation, see for instance
\cite{Gozzi:1984jx} and references therein. Such determinant depends on the
convention chosen for Heaviside theta function in zero: if $\theta\left(0\right)=0$
we are considering the Ito calculus while if $\theta\left(0\right)=1/2$
we are considering Stratonovich. 
In particular, as shown in \cite{Kupiainen:2006em}, using Ito discretization prescription 
the determinant does not enter in the path integral and the action has the same form of 
the action $S_\theta$ defined in equation (\ref{eq:def_Stheta}) below but with a different molecular diffusivity.
As we already said in this work we will follow Stratonovich discretization prescription  
and we shall
include the determinant contribution in the action via two Grassmann
odd fields which we refer to as ghosts. Such choice may seem somehow
more complicated since it doubles the number of fields. Nevertheless
we would like to keep such ghost fields since they carry a physical and geometrical
meaning in the path integral approach to classical mechanics \cite{Gozzi:1989bf,Gozzi:1993tm}
and make manifest the supersymmetry in stochastic processes, see \cite{Sakita:1986ad} and references therein. The ghost
action associated to the determinant is simply:
\begin{eqnarray*}
S_{gh} & = & \int\bar{c}\left(\partial_{t}+v^{i}\partial_{i}-\frac{\kappa}{2}\partial^{2}\right)c\,.
\end{eqnarray*}
The full action is thus composed of two parts: $S_{gh}$ and
\begin{eqnarray}
S_{\theta} & = & \int\left[\bar{\theta}\left(\partial_{t}+v^{i}\partial_{i}-\frac{\kappa}{2}\partial^{2}\right)\theta-\frac{1}{2}\bar{\theta}D_{\bar{\theta}}\bar{\theta}+\frac{1}{2}vD_{v}^{-1}v\right].  \label{eq:def_Stheta}  
\end{eqnarray}
Note that the field $\bar{\theta}$ is the response field present
in the Martin-Siggia-Rose formalism \cite{Martin:1973zz}, see also \cite{janssen1976,deDominicis1976},
and that $S_{\theta}$ and $S_{gh}$ are very similar due to the
fact that equation (\ref{eq:EOM_theta_force}) is linear in $\theta$. 
Thus the stochastic problem in equation (\ref{eq:EOM_theta_force}) has been mapped to the
field theory coming from expression (\ref{eq:Z_and_det}). Now we are able to exploit
the techniques available for standard field theory and in particular we will employ the
Functional Renormalization Group discussed in the next section.
Finally let us recall
the dimension of the various quantities entering in the action, we
have $\left[\theta\right]=T^{1/2}$, $\left[\bar{\theta}\right]=L^{-d}T^{-1/2}$,
$\left[D_{0}\right]=\frac{L^{2}}{T}L^{-\varepsilon}$ and $\left[\kappa\right]=\frac{L^{2}}{T}$.
Following \cite{Adzhemyan1998_incompressible} we define $g=D_{0}/\kappa$ which has dimensions
$\left[g\right]=L^{-\varepsilon}$.

\section{Functional Renormalization Group for the Kraichnan model \label{sec:Functional-Renormalization-Group}}

Functional Renormalization Group provides a non perturbative framework
for the implementation of the Wilsonian integration of high momentum
modes. For a general introduction we refer the reader to the following
list of reviews \cite{Morris:1998da,Bagnuls:2000ae,Berges:2000ew,Niedermaier:2006wt,Igarashi:2009tj}. We will work via the Effective Average
Action (EAA) which is a scale dependent generalization of the standard
effective action. One first define a modified generating functional
of connected Green's function $W_{k}$:
\begin{eqnarray*}
e^{W_{k}\left[J\right]} & = & \int{\cal D}\chi e^{-S-\Delta S_{k}+J\chi}
\end{eqnarray*}
where $\Delta S_{k}$ suppresses the integration of momentum modes
$p^{2}<k^{2}$ and is quadratic in the fields with a kernel $R_{k}$,
i.e.: $\Delta S_{k}=\frac{1}{2}\int\chi R_{k}\chi$. Note that this
acts like an infrared cutoff. Thus let $\tilde{\Gamma}_{k}$ be the
Legendre transform of $W_{k}$ and define the EAA subtracting
the cutoff action which we added at the beginning:
\begin{eqnarray*}
\Gamma_{k} & \equiv & \tilde{\Gamma}_{k}-\Delta S_{k}\,.
\end{eqnarray*}
The $k-$dependence of the functional $\Gamma_{k}$ satisfy the following
exact equation \cite{Wetterich:1992yh,Ellwanger:1993mw,Morris:1993qb}
\begin{eqnarray}
\partial_{t}\Gamma_{k} & = & \frac{1}{2}\mbox{Tr}\left[\left(\Gamma_{k}^{\left(2\right)}+R_{k}\right)^{-1}\partial_{t}R_{k}\right]\, \label{eq:flow_eq}
\end{eqnarray}
where $\partial_t = k\partial_k$ is the logarithmic derivative with respect to the cutoff
and $\Gamma_k^{\left(2\right)}$ is the Hessian of the EAA.
To concretely employ equation (\ref{eq:flow_eq}) we will need to resort
to some approximations and we will make an ansatz for $\Gamma_{k}$. Such
procedure has been proved robust in many fields, especially to determine
the scaling of statistical system at criticality, see \cite{Delamotte:2007pf} for
an overview.

Equation (\ref{eq:flow_eq}) will provide the running for the couplings
in our ansatz. Nevertheless we are also interested in the running
of composite operators. To derive the flow equation for composite
operators let us consider the one loop case. Besides the cutoff term
let us add to the action a source $\varepsilon$ coupled to a composite
operator $O$, then the one loop EAA reads
\begin{eqnarray*}
\Gamma_{k,1} & = & S +\varepsilon O+\frac{1}{2}\mbox{Tr}\log\left(S^{\left(2\right)}+R_{k}+\varepsilon O^{\left(2\right)}\right)
\end{eqnarray*}
where $S^{\left(2\right)}$ and $O^{\left(2\right)}$ are
the Hessians for the action and the composite operator respectively.
If we now derive the above expression with respect to the scale $k$
we obtain:
\begin{eqnarray*}
\partial_{t}\Gamma_{k,1} & = & \frac{1}{2}\mbox{Tr}\left(S^{\left(2\right)}+R_{k}+\varepsilon O^{\left(2\right)}\right)^{-1}\partial_{t}R_{k}\,.
\end{eqnarray*}
Finally, in order to derive the running of the composite operator,
we just need to take a functional derivative with respect to the source
$\varepsilon$ and set this to zero:
\begin{eqnarray*}
\partial_{t}O & = & -\frac{1}{2}\mbox{Tr}\left(S^{\left(2\right)}+R_{k}\right)^{-1}O^{\left(2\right)}\left(S^{\left(2\right)}+R_{k}\right)^{-1}\partial_{t}R_{k}\,.
\end{eqnarray*}
The exact flow equation for the running of the composite operator
can be found from the above expression just by ``RG-improving''
the one-loop case (exactly as it happens with the EAA). 
To see this explicitly we consider that, given a modified action $S+\varepsilon O$,
both the connected Green's function generating functional $W_k\left[ J, \varepsilon \right]$ and its Legendre 
transform $\tilde{\Gamma}_{k} \left[ \varphi , \varepsilon \right]$ depend on the source $\varepsilon$. 
In particular from the definition of $\tilde{\Gamma}_{k} $ and ${\Gamma}_{k} $ we have
\begin{equation*}
\Gamma_k \left[ \varphi , \varepsilon \right] = \int \varphi J - W_k\left[ J, \varepsilon \right] - \Delta S_k \left[ \varphi \right] \,.
\end{equation*}
In order to extract information regarding the composite operator $O$ we simply take a functional derivative with respect to $\varepsilon$
and observe that
\begin{equation*}
\frac{\delta}{\delta \varepsilon \left( x \right)} \Gamma_k \left[ \varphi , \varepsilon \right] \Bigr|_{\varepsilon =0} = - \frac{\delta W_k\left[ J, \varepsilon \right]}{\delta \varepsilon \left( x \right)}   \Bigr|_{\varepsilon =0} \,. 
\end{equation*}
This equation tells us that the running composite operator $\left[ O \right]_k $ is directly related
to the ${\Gamma}_{k} \left[ \varphi , \varepsilon \right]$ via a functional derivative with respect to $\varepsilon$. 
The flow equation for ${\Gamma}_{k} \left[ \varphi , \varepsilon \right]$ is simply given by equation (\ref{eq:flow_eq}) 
where the usual EAA is replaced with the $\varepsilon-$dependent EAA.
Thus the running of the first functional derivative of ${\Gamma}_{k} \left[ \varphi , \varepsilon \right]$ 
with respect to $\varepsilon$ can be directly obtained from (\ref{eq:flow_eq}) and, setting
$\varepsilon =0$ at the end, one obtains:
\begin{eqnarray}
\partial_{t}O_k & = & -\frac{1}{2}\mbox{Tr} \Bigr[ 
\left(\Gamma_{k}^{\left(2\right)}+R_{k}\right)^{-1}O_{k}^{\left(2\right)} \nonumber \\
& \, & \qquad \qquad \qquad \qquad \quad   
\left(\Gamma_{k}^{\left(2\right)}+R_{k}\right)^{-1}\partial_{t}R_{k} \Bigr] . \label{eq:flow_eq_composite_op}
\end{eqnarray}
Equation (\ref{eq:flow_eq_composite_op}) will give back a differential equation 
to which a suitable boundary condition is associated. Typically one defines
the composite operator of interest at some scale $\Lambda$. Clearly the RG evolution
generally produces mixing among different operators and the composite
operator has to be parametrized choosing a suitable ansatz.
For a detailed discussion on the flow equation for composite operators we refer to \cite{Igarashi:2009tj}.

Let us come back to the Kraichnan model. We choose the cutoff kernel
$R_{k}$ as
\begin{eqnarray}
R_{k} & = & \left(\begin{array}{ccccc}
0 & R_{k,\theta\bar{\theta}} & 0 & 0 & 0\\
R_{k,\bar{\theta}\theta} & 0 & 0 & 0 & 0\\
0 & 0 & R_{k,vv} & 0 & 0\\
0 & 0 & 0 & 0 & R_{k,c\bar{c}}\\
0 & 0 & 0 & R_{k,\bar{c}c} & 0
\end{array}\right)\,.\label{eq:cutoff_form_field_space}
\end{eqnarray}
The matrix acts on the following vector in field space: $\left(\theta,\bar{\theta},v,c,\bar{c}\right)$.
As common in the application of FRG to non-equilibrium systems we
have chosen an off-diagonal cutoff for the response field and its
companion (e.g.: $\bar{\theta}$ and $\theta$ respectively). Moreover
we do not cutoff the frequency since we will be able to perform integration
over frequency analytically. 
Note that adding the cutoff may break some of the symmetries analysed in
appendix \ref{sub:Symmetries}. Of course at the fixed point the scaling behaviour does
not depend on the form of the cutoff but breaking some symmetries
in the middle of the flow means that more terms have to be taken into
account. It is clear that if we consider
\begin{eqnarray}
R_{k} & = & \frac{1}{e^{\left(\frac{\Delta}{k^{2}}\right)^{\alpha}}-1} \left(\Delta \right)^{\alpha} \, , \label{eq:def_exp_cutoff}
\end{eqnarray}
where $\Delta=-\partial^2$, none of the symmetry is broken and in particular shift symmetry is
preserved. If we chose the optimized cutoff \cite{Litim:2001up}
\begin{eqnarray}
R_{k} & = & \left(k^{2\alpha}-\Delta^\alpha \right)\theta\left(k^{2}-\Delta \right)\label{eq:def_Litim_cutoff}
\end{eqnarray}
shift symmetry would be broken.

We are interested in the scaling of structure functions $S_{n}$ of
the field $\theta$ at equal time, i.e.:
\begin{eqnarray*}
S_{2n} & \equiv & \langle\left(\theta\left(t,x\right)-\theta\left(t,y\right)\right)^{2n}\rangle\,.
\end{eqnarray*}
Clearly $S_{2n}$ is given by a product of several different composite
operators and to determine the scaling behaviour of $S_{2n}$ is a
rather difficult task.  It is convenient to reason in terms of single
composite operators expressing $S_{2n}$ via the Operator Product
Expansion (OPE), or Short Distance Expansion. Such program works nicely
in the $\varepsilon-$expansion framework in \cite{Adzhemyan1998_incompressible} and we will follow closely
this procedure. Here the main idea is to identify the operators which
determine the scaling in such a way to be able to apply the FRG framework
afterwards. First of all let us consider the simplest case of $S_{2}$
and recall that the canonical dimension of the field is $\left[\theta\right]=T^{1/2}$.
For dimensional reason thus we have:
\begin{eqnarray}
S_{2} & = & T\cdot C_{1}\left(\mu_{T}T,\mu x\right)+C_{2}\left(\mu_{T}T,\mu x\right)\left[\theta^{2}\right] \nonumber \\
 &  & +T^{-1}C_{3}\left(\mu_{T}T,\mu x\right)\left[\theta^{4}\right]+\cdots +D_{1}\left(\mu_{T}T,\mu x\right)x^{2}\left[\partial\theta\partial\theta\right] \nonumber \\
 &  & +T^{-1} D_{2}\left(\mu_{T}T,\mu x\right)x^{4}\left[\left(\partial\theta\partial\theta\right)^{2}\right]+\cdots\,.\label{eq:generic_OPE_S2}
\end{eqnarray}
Note that odd terms in $\theta$ are not allowed by $\mathbb{Z}_{2}-$symmetry.
Moreover both the path integral and the correlation function in $S_{n}$
are shift invariant. As a consequence the operators entering in the
OPE are also shift invariant, i.e.: only derivatives of $\theta$
appear and all the terms in the first line, except the first one, vanish.
Finally let us recall that we are considering equal time correlators
and so $T$ must be set to zero. Now, if we make the hypothesis that
the time dependence of the various terms entering in (\ref{eq:generic_OPE_S2})
is not dramatically different from the canonical one, we can observe
that for $T\rightarrow0$ some terms vanish, some others stay finite
and some diverge. We assume that the divergent terms are eliminated
from the renormalization procedure so that only finite terms remain.
At this point we note that only operators like $\partial\theta\partial\theta$
are relevant in (\ref{eq:generic_OPE_S2}), i.e.: operators which
are not multiplied by overall factors of $T$. In full analogy one
can argue that for $S_{2n}$ the relevant composite operator is $\left(\partial\theta\partial\theta\right)^{2n}$.
This reasoning is at best heuristic and can be better justified
by diagrammatic considerations as presented in \cite{Adzhemyan1998_incompressible} where it is shown that  
only composite operators with at most $2n$ $\theta-$fields appear in
the OPE of $S_{2n}$. 
Here our main message is that the OPE suggests that the scaling of the structure
functions is encoded in the scaling of some particular composite operators
whose leading term in a derivative expansion is $\left(\partial\theta\partial\theta\right)^{2n}$.
Note that this is the operator you get from a simple Taylor expansion
of $S_{2n}$.

In order to derive the formula for the anomalous dimension of a composite
operator $O$ we consider the Callan-Symanzik equation and implement
renormalization multipling the operator by a coupling $Z_{O}$. Let
us consider the coupling appearing in $S_{\theta}$ and $S_{g}$,
the Callan-Symanzik equation reads: 
\begin{eqnarray}
\left(\mu\partial_{\mu}+\beta_{\kappa}\partial_{\kappa}+\gamma_{O}\right)\langle O\rangle & = & 0\label{eq:CallanSymanzik_O}
\end{eqnarray}
where we used the fact $D_{0}$ is not renormalized and thus does
not enter in the equation and $\gamma_{O}=\partial_{t}Z_{O}/Z_{O}$.
Moreover from dimensional analysis (in space and time) we have
\begin{eqnarray}
& & \left(-D_{0}\partial_{D_{0}}-\kappa\partial_{\kappa}+T\partial_{T}+d_{O}^{\omega}\right)\langle O\rangle  = 0\label{eq:T_dim_analysis_O}\\
& & \left(-\mu\partial_{\mu}+\left(2-\varepsilon\right)D_{0}\partial_{D_{0}} \right. \nonumber \\
& & \,\,\,\,\,\,\,\,\, \left. +2\kappa\partial_{\kappa}+x\partial_{x}-M\partial_{M}+d_{O}^{k}\right)\langle O\rangle =  0\label{eq:X_dim_analysis_O}
\end{eqnarray}
where $d_{O}^{\omega}$ and $d_{O}^{k}$ are the frequency and mass
dimensions of the operator $O$. Now let us plug (\ref{eq:CallanSymanzik_O})
in (\ref{eq:X_dim_analysis_O}), we obtain
\begin{eqnarray*}
& & \left(\beta_{\kappa}\partial_{\kappa}+\gamma_{O}+\left(2-\varepsilon\right)D_{0}\partial_{D_{0}} \right. \\
& & \,\,\,\,\,\,\,\,\, \left. +2\kappa\partial_{\kappa}+x\partial_{x}-M\partial_{M}+d_{O}^{k}\right)\langle O\rangle =  0\,.
\end{eqnarray*}
Now we insert (\ref{eq:T_dim_analysis_O}) in the above relation and find
\begin{eqnarray}
& & \left(\beta_{\kappa}\partial_{\kappa}+\gamma_{O}-\varepsilon D_{0}\partial_{D_{0}}+2T\partial_{T}  \right. \nonumber \\
& & \,\,\,\,\,\,\,\,\, \left. +2d_{O}^{\omega}+x\partial_{x}-M\partial_{M}+d_{O}^{k}\right)\langle O\rangle =  0\,.\label{eq:CSmanipulations_1}
\end{eqnarray}
Finally it is convenient to re-express the flow via the coupling $g=D_{0}/\kappa$
and to observe that we can rewrite the beta function of $\kappa$
via $g$ exploiting the fact that $D_{0}$ is not renormalized:
\begin{eqnarray*}
\beta_{\kappa}\partial_{\kappa} & = & \frac{\beta_{\kappa}}{\kappa}\kappa\partial_{\kappa}\\
 & = & \frac{k\frac{\partial}{\partial k}\left(\frac{\kappa}{D_{0}}\right)}{\frac{\kappa}{D_{0}}}\kappa\partial_{\kappa}=\frac{k\frac{\partial}{\partial k}\left(g^{-1}\right)}{g^{-1}}\kappa\partial_{\kappa}\\
 & = & -\frac{1}{g}\beta_{g}\kappa\partial_{\kappa}\,.
\end{eqnarray*}
Now we replace $\kappa\partial_{\kappa}$ using equation (\ref{eq:T_dim_analysis_O})
and we express $\beta_{g}$ via its dimensionless counterpart $\tilde{g}\equiv gk^{-\varepsilon}$.
Note that $\beta_{g}=\left(\varepsilon\tilde{g}+\beta_{\tilde{g}}\right)k^{\varepsilon}$
and so
\begin{eqnarray*}
\frac{1}{g}\beta_{g}\kappa\partial_{\kappa} & = & \frac{1}{\tilde{g}}\left(\varepsilon\tilde{g}+\beta_{\tilde{g}}\right)\left(-D_{0}\partial_{D_{0}}+T\partial_{T}+d_{O}^{\omega}\right)\\
 & = & \varepsilon\left(-D_{0}\partial_{D_{0}}+T\partial_{T}+d_{O}^{\omega}\right)
\end{eqnarray*}
at a non--Gaussian fixed point. Now we plug this in equation (\ref{eq:CSmanipulations_1}):
\begin{eqnarray}
& & \left(-\varepsilon\left(-D_{0}\partial_{D_{0}}+T\partial_{T}+d_{O}^{\omega}\right)+\gamma_{O}-\varepsilon D_{0}\partial_{D_{0}}+2T\partial_{T} \right. \nonumber \\
& & \,\,\,\,\,\,\,\,\, \left. +2d_{O}^{\omega}+x\partial_{x}-M\partial_{M}+d_{O}^{k}\right)\langle O\rangle = 0  \, ,\nonumber \\
& & \left(\left(2-\varepsilon\right)T\partial_{T}+\left(2-\varepsilon\right)d_{O}^{\omega} \right. \nonumber \\
& & \qquad \qquad  \left. +x\partial_{x}-M\partial_{M}+d_{O}^{k}+\gamma_{O}\right)\langle O\rangle = 0\,.\label{eq:scaling_composite_operator}
\end{eqnarray}
This equation will be used in section \ref{sec:Scaling-of-the-Sn} to derive the scaling of various
composite operators and to determine the scaling of the structure functions.
Summarizing we have introduced the FRG framework for the couplings
and for the composite operators. Thus we used the OPE to identify which
is the most important operator which contributes to the scaling of
the structure function.

\section{The IR fixed point \label{sec:The-IR-fixed}}

In this section we derive the IR fixed point whose scaling will be
studied in the next sections. The presence of an IR fixed point for
the Kraichnan model is known from $\varepsilon-$expansion
in \cite{Adzhemyan1998_incompressible}. In \cite{Adzhemyan1998_incompressible} it is argued that the beta function of
$g$, which is the coupling we choose to parametrize the RG flow,
is one loop exact. Thus we will first perform a one loop computation
within the FRG framework and afterwards we will comment on how the
fact that the beta function of $g$ is one loop exact can be understood
from an FRG perspective.

\begin{figure}
\begin{centering}
\includegraphics[scale=0.17]{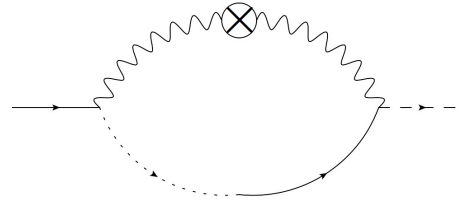} 
\par
\end{centering}
\caption{Diagrammatic representation of the flow equation (\ref{eq:flow_kappa_with_Gprop}).}
\label{img:kappa_Gvv_Gttb_Gvv}
\end{figure}

\begin{figure}
\begin{centering}
\includegraphics[scale=0.17]{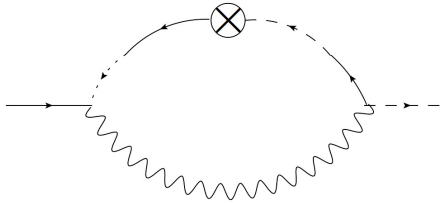}
\par
\end{centering}
\caption{Diagrammatic representation of the further contribution to the flow equation of $\kappa$ which vanish after frequency integration.}
\label{img:kappa_Gttb_Gvv_Gttb}
\end{figure}

We start by considering the ansatz:
\begin{eqnarray}
\Gamma_{k} & = & \int\left[\bar{\theta}\left(\partial_{t}+v^{i}\partial_{i}-\frac{\kappa}{2}\partial^{2}\right)\theta-\frac{1}{2}\bar{\theta}D_{\bar{\theta}}\bar{\theta}+\frac{1}{2}vD_{v}^{-1}v  \right. \nonumber \\
& &  \left. +\bar{c}\left(\partial_{t}+v^{i}\partial_{i}-\frac{\kappa}{2}\partial^{2}\right)c\right]\,.\label{eq:1loop_EAA_ansatz}
\end{eqnarray}
In order to evaluate the running of $\kappa$ it is convenient to consider
the flow equation for the functional derivatives of $\Gamma_{k}$.
In particular we derive the flow equation (\ref{eq:flow_eq}) with
respect to $\theta$ and $\bar{\theta}$ and extract the term proportional
to the momentum square. The central object in the flow equation is
the Hessian of the EAA which in field space and at zero field reads
\begin{eqnarray*}
\Gamma_{k}^{\left(2\right)} & = & \left(\begin{array}{ccccc}
0 & \gamma_{\theta\bar{\theta}} & 0 & 0 & 0\\
\gamma_{\bar{\theta}\theta} & \gamma_{\bar{\theta}\bar{\theta}} & 0 & 0 & 0\\
0 & 0 & \gamma_{vv} & 0 & 0\\
0 & 0 & 0 & 0 & -\gamma_{c\bar{c}}\\
0 & 0 & 0 & \gamma_{\bar{c}c} & 0
\end{array}\right)\,.
\end{eqnarray*}
Thus the regularized propagator $G_{k}=\left(\Gamma_{k}^{\left(2\right)}+R_{k}\right)^{-1}$
has the form
\begin{eqnarray*}
G_{k} & = & \left(\begin{array}{ccccc}
G_{\bar{\theta}\bar{\theta}} & G_{\theta\bar{\theta}} & 0 & 0 & 0\\
G_{\bar{\theta}\theta} & 0 & 0 & 0 & 0\\
0 & 0 & G_{vv} & 0 & 0\\
0 & 0 & 0 & 0 & G_{c\bar{c}}\\
0 & 0 & 0 & -G_{\bar{c}c} & 0
\end{array}\right)\,.
\end{eqnarray*}
In particular we have
\begin{eqnarray*}
G_{\theta\bar{\theta}} & = & G_{\bar{\theta}\theta}^{*}\\
 & = & \frac{1}{i\omega+\frac{\kappa}{2}q^{2}+R_{k}}\\
G_{vv} & = & \frac{D_{0}}{\left(q^{2}\right)^{\frac{d+\varepsilon}{2}}+R_{k}}P_{ij}\\
G_{c\bar{c}} & = & G_{\bar{c}c}^{*}\\
 & = & \frac{1}{i\omega+\frac{\kappa}{2}q^{2}+R_{k}}\,.
\end{eqnarray*}
Schematically the flow equation for two point function has the following
form:
\begin{eqnarray}
\partial_{t}\frac{\delta^{2}\Gamma_{k}}{\delta\theta\delta\bar{\theta}} & = & \frac{1}{2}\mbox{Tr}\Bigr[G_{k}\cdot\frac{\delta\Gamma_{k}^{\left(2\right)}}{\delta\theta}\cdot G_{k}\cdot\frac{\delta\Gamma_{k}^{\left(2\right)}}{\delta\bar{\theta}}\cdot G_{k}\cdot\partial_{t}R_{k}
\nonumber \\
 &  &
 +G_{k}\cdot\frac{\delta\Gamma_{k}^{\left(2\right)}}{\delta\bar{\theta}}\cdot G_{k}\cdot\frac{\delta\Gamma_{k}^{\left(2\right)}}{\delta\theta}\cdot G_{k}\cdot\partial_{t}R_{k} \nonumber \\
& &  -G_{k}\cdot\frac{\delta^{2}\Gamma_{k}^{\left(2\right)}}{\delta\theta\delta\bar{\theta}}\cdot G_{k}\cdot\partial_{t}R_{k}\Bigr]\,.\label{eq:generic_2point_flow_eq}
\end{eqnarray}
Note that the term in the third line does not contribute since there
is no such vertex in the ansatz (\ref{eq:1loop_EAA_ansatz}). Thus
we find that
\begin{eqnarray}
\partial_{t}\frac{\delta^{2}\Gamma_{k}}{\delta\theta\delta\bar{\theta}} & = & G_{vv}\gamma^{\left(3\right)}G_{\theta\bar{\theta}}\gamma^{\left(3\right)}G_{vv}\partial_{t}R_{k,vv}\label{eq:flow_kappa_with_Gprop}
\end{eqnarray}
where $\gamma^{\left(3\right)}$ denotes the vertex coming from the
term $\bar{\theta}v^{i}\partial_{i}\theta$ in the action. 
Equation (\ref{eq:flow_kappa_with_Gprop}) can be written in a diagrammatic
fashion as shown in Figure \ref{img:kappa_Gvv_Gttb_Gvv}.
Note that there are further contributions to the flow equation for
the two point function but they vanish upon integration over the frequency
(see appendix \ref{sec:Appendix:-some-frequency} for some details
on the frequency integration). These diagrams are shown in Figure
\ref{img:kappa_Gttb_Gvv_Gttb}.
The final result is:
\begin{eqnarray}
& & G_{vv}\gamma^{\left(3\right)}G_{\theta\bar{\theta}}\gamma^{\left(3\right)}G_{vv}\partial_{t}R_{k,vv}  =  -\frac{1}{2}\left(\frac{1}{\left(2\pi\right)^{d}}\frac{2\pi^{d/2}}{\Gamma\left(\frac{d}{2}\right)}\frac{d-1}{d}\right)\frac{D_{0}}{2} \nonumber \\
& & \qquad \qquad \qquad
\cdot\int_{0}^{\infty}dz\frac{z^{d/2-1}}{\left(z^{\frac{d+\varepsilon}{2}}+R_{k,vv}\right)^{2}}\partial_{t}R_{k,vv}\,.\label{eq:flow_eq_kappa_dim_int}
\end{eqnarray}
If we use the exponential cutoff (\ref{eq:def_exp_cutoff}) in the
above equation we obtain
\begin{eqnarray}
& & G_{vv}\gamma^{\left(3\right)}G_{\theta\bar{\theta}}\gamma^{\left(3\right)}G_{vv}\partial_{t}R_{k,vv} = -\frac{1}{2}\left(\frac{1}{\left(2\pi\right)^{d}}\frac{2\pi^{d/2}}{\Gamma\left(\frac{d}{2}\right)}\frac{d-1}{d}\right) \nonumber \\
& & \qquad \qquad \qquad \qquad 
\frac{D_{0}}{2}k^{-\varepsilon}\cdot\left(2\frac{d+\varepsilon}{d}\Gamma\left[1+\frac{d}{d+\varepsilon}\right]\right)\label{eq:rhs_kappa_exp_cutoff}
\end{eqnarray}
and for the optimized cutoff (\ref{eq:def_Litim_cutoff})
\begin{eqnarray}
& & G_{vv}\gamma^{\left(3\right)}G_{\theta\bar{\theta}}\gamma^{\left(3\right)}G_{vv}\partial_{t}R_{k,vv} = -\frac{1}{2}\left(\frac{1}{\left(2\pi\right)^{d}}\frac{2\pi^{d/2}}{\Gamma\left(\frac{d}{2}\right)}\frac{d-1}{d}\right)  \nonumber \\
& & \qquad \qquad \qquad \qquad  \qquad \qquad  \qquad  
\frac{D_{0}}{2}k^{-\varepsilon}\cdot\left(2\frac{d+\varepsilon}{d}\right)\,.\label{eq:rhs_kappa_litim_cutoff}
\end{eqnarray}
The fact that beta functions coming from
(\ref{eq:rhs_kappa_exp_cutoff}) and (\ref{eq:rhs_kappa_litim_cutoff})
are slightly different has to be expected since the beta functions
are scheme dependent and using different cutoffs corresponds to a
change of scheme in the FRG language. Scheme independence has to be
recovered in physical quantities like the critical exponents as we will 
check in a moment. In this
case it is convenient to parametrize the flow via the coupling $g=D_{0}/\kappa$
and to introduce its dimensionless analogue $\tilde{g}=gk^{-\varepsilon}$.
Note that the flow equation (\ref{eq:flow_eq_kappa_dim_int}) has
the following generic form
\begin{eqnarray*}
\partial_{t}\left(\frac{\kappa}{2}\right) & = & -AD_{0}k^{-\varepsilon}
\end{eqnarray*}
where $A$ is a constant which depend on the dimension and on $\varepsilon$.
In term of the dimensionless coupling $\tilde{g}$ the above equation
reads:
\begin{eqnarray*}
\partial_{t}\tilde{g} & = & -\varepsilon\tilde{g}+2A\tilde{g}^{2}\,.
\end{eqnarray*}
The IR fixed point thus reads $\tilde{g}_{*}=\varepsilon/\left(\text{2}A\right)$
and the critical exponent is $\beta_{\tilde{g}}^{\prime}=\varepsilon$
independently of the cutoff. 

Let us compare our procedure with the one adopted in \cite{Adzhemyan1998_incompressible}. In
\cite{Adzhemyan1998_incompressible} the authors set up a field theoretic framework analogous
to the one we have seen in section \ref{sec:Path-integral-formulation}. The only difference being the
fact that the ghost fields are not present. The one loop correction
to the molecular diffusivity is computed and the diagram is UV convergent.
However when $\varepsilon \rightarrow 0$ a divergence appears.
In full analogy with dimensional regularization the authors perform
an $\varepsilon-$expansion and subtract the poles. This leads to
a prescription for the RG flow which can be used to obtain a resummed
result. When one performs this $\varepsilon-$expansion one notes
that the scale $m$ disappears from the leading expression proportional
to $1/\varepsilon$ and disappears from the flow equations. In our
case we simply set $m=0$ from the very beginning. 

Note that in comparison to other works in the literature there are two main differences.
First the velocity covariance (\ref{eq:def_Dv_expl}) adopted in this work and in \cite{Adzhemyan1998_incompressible} 
differs from the one used in other works since
the coefficient $D_0$ in (\ref{eq:def_Dv_expl}) is replaced with $D_0 \varepsilon$  (see for instance \cite{Gawedzki:1995zz,Bernard:1996um,BernardGawedzkiKupiainen,Kupiainen:2006em}).
Moreover the problem can be represented using the Ito discretization prescription
in place of the Stratonovich one.
With this definition the velocity covariance is finite in the limit $\varepsilon \rightarrow 0$
and the previously mentioned divergence 
for $\varepsilon \rightarrow 0$ is not present \cite{Kupiainen:2006em}. 
In this case the RG flow
of the action is trivial and non-trivial flows can be found for composite fields
like $\left( \partial \theta \right)^n$. 
We refer to \cite{Kupiainen:2006em} for more details
on the use of Ito prescription and the RG flows found in this case.

Finally in \cite{Adzhemyan1998_incompressible}
it is argued that the beta function of $\tilde{g}$ is one-loop exact.
To understand how this should be interpreted in the FRG scheme let
us consider again the ansatz (\ref{eq:1loop_EAA_ansatz}) and plug
this in the flow equation for the two point function. We want to understand
if further terms are generated by the RG flow or not. It is immediate
to note that no further term is generated: the ``trace factor''
found by contracting the vertices with the propagator
is $-P^{2}\left(1-X^{2}\right)$ and the propagator $G_{\theta\bar{\theta}}$
(which depends on $p_{i}+q_{i}$) loses its $P-$dependence after
the frequency integration. As a consequence the only $P-$dependence
comes from the trace factor. This means that the l.h.s. and r.h.s.
of (\ref{eq:flow_kappa_with_Gprop}) are proportional to $P^{2}$,
a very uncommon fact in FRG computations. One can be even more general
and start from the following ansatz
\begin{eqnarray*}
\Gamma_{k} & = & \int\left[\bar{\theta}\left(\partial_{t}+v^{i}\partial_{i}+\frac{\kappa}{2}f\left(-\partial^{2}\right)\right)\theta-\frac{1}{2}\bar{\theta}D_{\bar{\theta}}\bar{\theta}+\frac{1}{2}vD_{v}^{-1}v \right. \\
& & \left. +\bar{c}\left(\partial_{t}+v^{i}\partial_{i}+\frac{\kappa}{2}f\left(-\partial^{2}\right)\right)c\right]\,.
\end{eqnarray*}
Repeating the same steps the flow equation itself tells that $f\left(-\partial^{2}\right)=-\partial^{2}$.
We interpret this as an argument telling us that the beta function
of $g$ is one-loop exact. Clearly one may consider even more general
ansatzs which are consistent with all the symmetries we analyzed in
appendix \ref{sub:Symmetries}. For example one can add higher powers
in the fields via terms like:
\[
\bar{\theta}^{2}f\left(\nabla_{t},-\partial^{2}\right)\theta^{2}+2\bar{\theta}\bar{c}f\left(\nabla_{t},-\partial^{2}\right)\theta c\,.
\]
We will not consider such terms but these terms should be taken into account in
a more complete computation.

\section{RG flow of composite operators \label{sec:RG-flow-of_composite_operators}}

In this section we study the scaling of the structure function at
the IR fixed point found in section \ref{sec:The-IR-fixed}. We will
first study the renormalization of the operator $\left(\partial\theta\partial\theta\right)^{n}$
and then we will move to consider $\theta^{m}$. The results found
in this section will be used in section \ref{sec:Scaling-of-the-Sn}
to obtain the correction to the scaling of $S_{2n}$.

\subsection{The operator $\left(\partial\theta\partial\theta\right)^{n}$ \label{sub:The-operator-dTheta2_n}}

In section \ref{sec:Functional-Renormalization-Group} we argued that
the scaling of the structure functions $S_{2n}$ is encoded in the
scaling of the composite operator $\left(\partial\theta\partial\theta\right)^{n}$.
The argument is based on the OPE of $S_{2n}$. Let us stress that
if we take into account only the operator $\left(\partial\theta\partial\theta\right)^{n}$
we are making an approximation since other operators also contribute,
e.g.: $\left(\partial^{2}\theta\partial^{2}\theta\right)^{n}$. In
this sense we apply the logic of the derivative expansion to the composite
operator coming from the OPE of $S_{2n}$.

From now on we focus on the RG flow of the operator $O=\left(\partial\theta\partial\theta\right)^{n}$.
To compute its running we introduce a coupling $Z_{O}$ which multiplies
$O$ and use equation (\ref{eq:flow_eq_composite_op}). To extract
the running of $Z_{O}$ we perform $2n$ functional derivatives with
respect to $\theta$ and set all the fields to zero. Thus l.h.s. is
proportional to $P^{2n}$ and has to be compared with the terms also
proportional to $P^{2n}$ obtained by taking $2n$ functional derivatives
of the r.h.s. of (\ref{eq:flow_eq_composite_op}). Note that the Hessian of the operator $O$
appears in equation (\ref{eq:flow_eq_composite_op}) and $2n-2$ functional derivatives will act on it while the
remaining two will act on the regularized propagators of the EAA. Moreover
the l.h.s. of the equation is proportional to $\partial_{t}Z_{O}$
while the r.h.s. to $Z_{O}$ so it will be easy to extract the anomalous
dimension $\partial_{t}Z_{O}/Z_{O}$. The flow equation obtained in
this way can be represented diagrammatically as in Figures \ref{img:OP_Gvv_Gttb_Gttb_Gvv}, \ref{img:OP_Gttb_Gttb_Gvv_Gttb} 
and \ref{img:OP_Gttb_Gvv_Gttb_Gttb} where the black dot denotes the insertion of the Hessian of the
composite operator with $2n-2$ functional derivatives attached. 
\begin{figure}
\begin{centering}
\includegraphics[scale=0.17]{OP_Gvv_Gttb_Gttb_Gvv.jpeg}
\par
\end{centering}
\caption{Diagrammatic representation of a contribution to flow equation for composite operators.}
\label{img:OP_Gvv_Gttb_Gttb_Gvv}
\end{figure}
\begin{figure}
\begin{centering}
\includegraphics[scale=0.17]{OP_Gttb_Gttb_Gvv_Gttb.jpeg}
\par
\end{centering}
\caption{Diagrammatic representation of a contribution to flow equation for composite operators.}
\label{img:OP_Gttb_Gttb_Gvv_Gttb}
\end{figure}
\begin{figure}
\begin{centering}
\includegraphics[scale=0.17]{OP_Gttb_Gvv_Gttb_Gttb.jpeg}
\par
\end{centering}
\caption{Diagrammatic representation of a contribution to flow equation for composite operators.}
\label{img:OP_Gttb_Gvv_Gttb_Gttb}
\end{figure}
We limit ourselves to consider the leading term in the $\varepsilon-$expansion
of the structure function which amounts to a one loop computation.
Our main objective is
to test the tools introduced so far and reproduce the known leading
correction.
In each diagram the ``trace factor'' between the vertices and the
velocity projector gives $P^{2}\left(1-X^{2}\right)$, where $X$ is the
cosine between the external and the loop momentum. The Hessian of
the composite operator reads $O^{\left(2\right)}=2nP^{2\left(n-1\right)}Q^{2}\left[1+X^{2}2\left(n-1\right)\right]$,
where $P$ and $Q$ are the modulus of the external and loop momentum
respectively. In each diagram the product of the ``trace factor''
and of $O^{\left(2\right)}$ gives a term of order $P^{2n}$. Thus
neglecting all the $P-$dependence in the regularized propagators
we obtain the running of $Z_{O}$ and the angular integral factors
out. The contribution to $\partial_{t}Z_{O}/Z_{O}$ from the diagram
of Figure \ref{img:OP_Gvv_Gttb_Gttb_Gvv} is
\begin{eqnarray*}
& &  -\frac{2^{-d}\pi^{-d/2}}{\Gamma\left(\frac{d}{2}\right)}   \frac{d-1}{d}
\left(n\frac{d+2n}{d+2}\right)\frac{D_{0}}{\kappa}   \\
& & \int_{0}^{\infty}dz\, z^{d/2}\frac{1}{z+R_{k,\theta\bar{\theta}}}\frac{1}{\left(z^{\frac{d+\varepsilon}{2}}+R_{k,vv}\right)^{2}}\partial_{t}R_{k,vv}\,.
\end{eqnarray*}
Furthermore both the diagrams of Figures \ref{img:OP_Gttb_Gttb_Gvv_Gttb} and \ref{img:OP_Gttb_Gvv_Gttb_Gttb} give the following
contribution to $\partial_{t}Z_{O}/Z_{O}$: 
\begin{eqnarray*}
& & - \frac{2^{-d}\pi^{-d/2}}{\Gamma\left(\frac{d}{2}\right)}  \frac{d-1}{d}
 \left(n\frac{d+2n}{d+2}\right)\frac{1}{2}\frac{D_{0}}{\kappa}   \\
& & \int_{0}^{\infty}dz\, z^{d/2}\frac{1}{z^{\frac{d+\varepsilon}{2}}+R_{k,vv}}\frac{1}{\left(z+R_{k,\theta\bar{\theta}}\right)^{2}}\partial_{t}R_{k,\theta\bar{\theta}}\,.
\end{eqnarray*}
The integrals can be solved analytically employing the optimized cutoff
(\ref{eq:def_Litim_cutoff}) and, summing all the contributions, the
leading contribution for small $\varepsilon$ is
\begin{eqnarray*}
\frac{\partial_{t}Z_{O}}{Z_{O}} & = & - \frac{2^{1-d}\pi^{-d/2}}{\Gamma\left(\frac{d}{2}\right)}\frac{d-1}{d}
\left(n\frac{d+2n}{d+2}\right)\tilde{g} \,.
\end{eqnarray*}
At the leading order in $\varepsilon$ we have:
\begin{eqnarray*}
\tilde{g}_{*} & = & \frac{d}{d-1}\left[\frac{1}{\left(2\pi \right)^d}\left(\frac{2\pi^{d/2}}{\Gamma\left(\frac{d}{2}\right)}\right)\right]^{-1}\varepsilon\,.
\end{eqnarray*}
Therefore we obtain
\begin{eqnarray*}
\frac{\partial_{t}Z_{O}}{Z_{O}}\Bigr|_{\tilde{g}_{*}} & = & -\left[n\frac{d+2n}{d+2}\right]\varepsilon\,.
\end{eqnarray*}
Although this computation is shown using the optimized cutoff we checked
that the fixed point anomalous dimension above is very stable under
changes of the cutoff kernel $R_{k}$.

Let us note that differently from what we have seen in section \ref{sec:The-IR-fixed}
the r.h.s. of the flow equation for composite operators generates new terms.
Indeed it is possible to verify that terms of higher power in $P$ are generated expanding
those propagators which are function of $ \left( p_i+q_i \right)$.
This means that, within our approximation, the system is not closed. This is a common
problem in the FRG approach and the crucial point is to parametrize the quantities of interest
with sufficiently many terms. Here, retaining just one operator, we did a rather crude approximation and  
this precludes for instance the possibility of extracting correct values of the anomalous dimension
for large values of $\varepsilon$.
However we stress that hopefully such difficulties can be solved by employing larger truncations
as usually happens in the FRG applications. In order to understand if this methodology is efficient
for the Kraichnan model
one should increase the truncation and check the stability of the results.
This however is beyond the scope of the present work.

\subsection{The operator $\theta^{m}$ \label{sub:The-operator-theta_m}}

Besides the anomalous dimension of the operator $\left(\partial\theta\partial\theta\right)^{n}$
it is useful to consider also the anomalous dimension of operators
$\theta^{m}$. It turns out that the anomalous of $\theta^{m}$ dimension
is zero. To see this we consider the flow equation for the composite
operator $\theta^{m}$ which is fully analogous to the one for $\left(\partial\theta\partial\theta\right)^{n}$.
One only needs to change the ``composite operator vertex'' in the
diagrams. This vertex is simply given by $m\left(m-1\right)$ (from the Hessian
of $\theta^{m}$). Now we recall that the trace over the vertices
and the projectors in these diagrams gives an overall factor of $P^{2}$.
Since the definition of the running of $Z_{\theta^{m}}$ is taken
at zero momentum we see that the r.h.s. of the flow equation for $Z_{\theta^{m}}$
is zero.

As an aside comment let us note that a non--trivial result is found
considering the Kraichnan model for a compressible fluid as done in
\cite{Adzhemyan1998_compressible}. In this case the action reads
\begin{eqnarray*}
S_{\theta} & = & \int\left[\bar{\theta}\left(\partial_{t}\theta+\partial_{i}\left(v^{i}\theta\right)-\frac{\kappa}{2}\partial^{2}\theta\right)-\frac{1}{2}\bar{\theta}D_{\bar{\theta}}\bar{\theta}+\frac{1}{2}vD_{v}^{-1}v\right]
\end{eqnarray*}
with
\begin{eqnarray*}
D_{v} & = & \frac{D_{0}}{\left(2\pi\right)^{d}}\delta\left(t-t^{\prime}\right)\int d^{d}k\,\frac{P_{ij}^{T}+\alpha P_{ij}^{L}}{\left(k^{2}+m^{2}\right)^{\frac{d}{2}+\frac{\varepsilon}{2}}}e^{ikx_{1}-ikx_{2}} \, ,
\end{eqnarray*}
where $P_{ij}^{T}$ and $P_{ij}^{L}$ are the transverse and longitudinal
projectors respectively. This model has been studied in \cite{Adzhemyan1998_compressible}
and we have checked at one loop that the fixed point and the anomalous
dimensions $\partial_{t}Z_{\theta^{m}}/Z_{\theta^{m}}$ are reproduced
by our framework.

\section{Scaling of the structure functions \label{sec:Scaling-of-the-Sn}}

We use the results of sections \ref{sec:The-IR-fixed} and \ref{sec:RG-flow-of_composite_operators}
in order to derive the running of the structure functions. To achieve this we will consider the OPE of $S_{2n}$ \cite{Adzhemyan1998_incompressible}.
Let us note that
\begin{eqnarray*}
S_{2n} & = & \langle\left(\theta\left(x\right)-\theta\left(y\right)\right)^{2n}\rangle\\
 & = & \sum_{k=0}^{2n}\left(\begin{array}{c}
2n\\
k
\end{array}\right)\langle\theta\left(x\right)^{2n-k}\theta\left(y\right)^{k}\rangle\,.
\end{eqnarray*}
Repeating the same steps which lead to equation (\ref{eq:scaling_composite_operator})
we obtain that all the terms in the above sum satisfy the following
equation
\begin{eqnarray*}
\left[r\partial_{r}+\left(2-\varepsilon\right)T\partial_{T}-n\left(2-\varepsilon\right)-M\partial_{M}\right]\langle\theta{}^{2n-k}\theta{}^{k}\rangle & = & 0
\end{eqnarray*}
where $r=\left|x-y\right|$. This tells us that
\begin{eqnarray*}
S_{2n} & \sim & r^{n\left(2-\varepsilon\right)}f\left(Mr\right)\,.
\end{eqnarray*}
In order to understand the $M-$dependence in $S_{2n}$ we consider
the OPE. In section \ref{sec:Functional-Renormalization-Group} we
argued that the leading term in the OPE of $S_{2n}$ is given by $\left(\partial\theta\partial\theta\right)^{n}$
thus 
\begin{eqnarray*}
S_{2n} & \approx & \langle C\left(r\right)\left(\partial\theta\partial\theta\right)^{n}\rangle
\end{eqnarray*}
where $C\left(r\right)$ is the first Wilson coefficient of the expansion,
a function of $\left(x-y\right)/2$, and $\left(\partial\theta\partial\theta\right)^{n}$
is a composite operator in $\left(x+y\right)/2$. Thus we have
\begin{eqnarray*}
S_{2n} & \approx & C\left(r\right)\langle\left(\partial\theta\partial\theta\right)^{n}\rangle
\end{eqnarray*}
where $\langle\left(\partial\theta\partial\theta\right)^{n}\rangle$
is a constant in the sense that it does not dependent on any coordinates.
Now we consider equation (\ref{eq:scaling_composite_operator}) for
the above correlation and deduce that
\begin{eqnarray*}
\left(-M\partial_{M}+n\varepsilon+\gamma_{O}^{*}\right)\langle\left(\partial\theta\partial\theta\right)^{n}\rangle & = & 0
\end{eqnarray*}
where we used the fact that $\langle\left(\partial\theta\partial\theta\right)^{n}\rangle$
does not depend on any coordinates. This tells us that
\begin{eqnarray}
\langle\left(\partial\theta\partial\theta\right)^{n}\rangle & \sim & M^{n\varepsilon+\gamma_{O}^{*}}\,.\label{eq:M_dependence_S2n}
\end{eqnarray}
Finally we suppose that the Wilson coefficient $C\left(r\right)$
does not depend on $M$ in the limit $M\rightarrow0$, i.e.: that
$C\left(r\right)$ is regular in $M$. Note that the limit $M\rightarrow0$,
i.e.: $L_{F}\gg r$, is justified by the fact that we are studying
the inertial regime of the system. This tells us that the whole $M-$dependence
in $S_{2n}$ is given by expression (\ref{eq:M_dependence_S2n}).
Therefore we conclude that
\begin{eqnarray}
S_{2n} & \sim & r^{n\left(2-\varepsilon\right)}\left(Mr\right)^{n\varepsilon+\gamma_{O}^{*}}\label{eq:scaling_S2n_rMr}
\end{eqnarray}
where
\begin{eqnarray*}
n\varepsilon+\gamma_{O}^{*} & = & -\frac{2n\left(n-1\right)}{d+2}\varepsilon\,.
\end{eqnarray*}
This is the leading correction found in \cite{Gawedzki:1995zz,Chertkov:1995wi} via the zero mode
approach and in \cite{Adzhemyan1998_incompressible} via the $\varepsilon-$expansion.

Let us note that we can derive the scaling of $S_{2n}$ also via the
following reasoning. Since $\langle\left(\partial\theta\partial\theta\right)^{n}\rangle$
does not depend on $r$ we can limit ourselves to study the $r-$dependence
in $S_{2n}$ studying the Wilson coefficient $C\left(r\right)$. We
suppose that $C\left(r\right)$ is regular in $M$ and consider the
$M\rightarrow0$ limit. The Callan-Symanzik equation for the Wilson
coefficient is:
\begin{eqnarray*}
\left(\mu\partial_{\mu}+\beta_{g}\partial_{g}-\gamma_{O}\right)C & = & 0
\end{eqnarray*}
where we used the fact that the anomalous dimension of the composite
operators $\theta^{2n-k}$ and $\theta^{k}$ are zero. Once again
we repeat the steps which lead to equation (\ref{eq:scaling_composite_operator})
and obtain
\begin{eqnarray*}
\left(r\partial_{r}-2n+\left(2-\varepsilon\right)T\partial_{T}-\gamma_{O}^{*}-M\partial_{M}\right)C & = & 0\,.
\end{eqnarray*}
Thus we obtain
\begin{eqnarray*}
C\left(r\right) & \sim & r^{2n+\gamma_{O}^{*}}
\end{eqnarray*}
where the exponent is exactly the one coming from the sum of the exponents
in (\ref{eq:scaling_S2n_rMr}).

\section{Summary and outlooks}

We have explored the application of the FRG framework to the Kraichnan
model. In particular we studied the scaling of the structure functions
in the Kraichnan model and reproduced the leading corrections by means
of FRG techniques. From the field theoretical point of view the structure
functions are correlations of composite operators. Thus in section
\ref{sec:Functional-Renormalization-Group} we recalled how to compute
the running of composite operators using the flow equation for the
EAA. In section \ref{sec:The-IR-fixed} and \ref{sec:RG-flow-of_composite_operators}
we studied the IR fixed point of the model and the anomalous dimensions
of various composite operators at the fixed point. The scaling of the
structure functions in the Kraichnan model has been found in section
\ref{sec:Scaling-of-the-Sn} using the OPE. 

The path integral formulation adopted in this work has been explained
in section \ref{sec:Path-integral-formulation} while the associated symmetries 
and some non--renormalization theorems are discussed in appendix \ref{sub:Symmetries}.
In the path integral
formulation of the problem a determinant naturally arises and we kept
track of this by expressing the determinant via the integration over
some ghost fields. The ghost sector carries two new symmetries which
we called BRS symmetry (mixing $\theta$ and $\bar{c}$) and ghost
symmetry. These symmetries are very similar to the ones present in
the path integral formulation for classical mechanics \cite{Gozzi:1989bf}. In
the latter case more symmetries are present, this is due to the fact
that the equations of motion for the Kraichnan model are not Hamiltonian.
Let us note that the Navier--Stokes equation can be formulated as Hamiltonian
equations of motion setting the viscosity to zero and using the so
called Clebsh variables. It may turn out interesting to study the
role of the symmetries and of the RG flow in this context. Moreover,
as far as classical mechanics is concerned, the ghost fields carry
information on the Lyapunov exponents of the system and thus on the
chaoticity/stochasticity of the system \cite{Gozzi:1993tm} and we feel that
they could play a role also in the path integral approach to turbulence.
Here we limit ourselves to note that in the Kraichnan model the equation of motion
for the ghost $c$ corresponds to unforced solutions which are the ones
involved in the zero mode approach. We hope to study
these topics in the future.

Finally we would like to stress that the techniques adopted in this
work can be applied to more complicated systems, for instance the
randomly stirred Navier--Stokes equation. On general grounds this
will require to fully exploit the functional character of the flow
equation in order to be able to account for an infinite dimensional
ansatz and not to rely on the derivative expansion. This has been
achieved successfully in the case of the Kardar-Parisi-Zhang equation
in \cite{Canet:2011ez}. In the Kraichnan model the situation is somehow simpler
since the IR fixed point is perturbative for small $\varepsilon$
and the beta function of $g$ is one loop exact \cite{Adzhemyan1998_incompressible}. Nevertheless
we believe that the logic adopted here, especially with regards to
composite operators, can be used in other contexts and more work is
needed in this direction. The advantage of the FRG framework is that
it does not rely on a perturbative expansion and that genuinely
non--perturbative results can be obtained provided that the EAA is
parametrized with sufficient accuracy. 
In particular we hope that a similar framework to the one adopted in this
work may be applied to more complicated situations, like the Navier--Stokes equation. 

\begin{acknowledgments}
The author acknowledges the support of the Foundation Blanceflor Boncompagni Ludovisi, n\'{e}e Bildt. 
The author is grateful to Martin Reuter for helpful discussions.
\end{acknowledgments}


\appendix

\section{symmetries \label{sub:Symmetries}}

In this appendix we consider the symmetries associated to the Krainchnan
model. We first consider the symmetries associated to the action $S_{\theta}$
which can be easily extended to the ghost action. Finally we consider
symmetries coming from non trivial cancellation in $S_{\theta}+S_{gh}$.
From these symmetries we will be able to derive some non-renormalization
theorems. Such program has been carried out in great detail for the
Kardar-Parisi-Zhang equation in \cite{Canet:2011ez} and more recently for the
Navier-Stokes equation in \cite{Canet:2014cta}. In these cases the analysis
showed that symmetries entail important constraints on the RG flow.
Here we perform a similar analysis also including the ghost sector.

The action is symmetric under a sort of $\mathbb{Z}_{2}$ transformation:
$\theta\rightarrow-\theta$ and $\bar{\theta}\rightarrow-\bar{\theta}$.
This tells us that there is an even number of $\theta$ and $\bar{\theta}$
fields in any monomial of the effective action. Moreover it is easy
to verify the invariance under constant shift in $\theta$
\begin{eqnarray*}
\theta & \rightarrow & \theta+u\,.
\end{eqnarray*}
Clearly this property is enjoyed also by the effective action. In
analogy with what has been done in \cite{Canet:2011ez} we consider a time dependent
shift:
\begin{eqnarray*}
\theta & \rightarrow & \theta+u\left(t\right)\,.
\end{eqnarray*}
This is not a symmetry but the bare action is non-invariant just by
a term which is linear in field space. In particular the only time
derivative appears in
\begin{eqnarray*}
\bar{\theta}\partial_{t}\theta & \rightarrow & \bar{\theta}\partial_{t}\theta+\bar{\theta}\partial_{t}u\left(t\right)\,.
\end{eqnarray*}
Since the non-invariance of the action is given just by a term which
is linear in the field we can derive the following Ward Idendity (WI):%
\footnote{We use the following conventions: $\exp W=\int\exp\left[-S+J \cdot \phi\right]$
and $\Gamma=\int J\varphi-W$. Thus $J=\delta_{\varphi}\Gamma$.%
}
\begin{eqnarray*}
0 & = & \langle J\cdot u\left(t\right)-\int\bar{\theta}\partial_{t}u\left(t\right)\rangle\\
 & = & \langle\left(\frac{\delta\Gamma}{\delta\theta}\cdot u\left(t\right)-\int\bar{\theta}\partial_{t}u\left(t\right)\right)\rangle\,.
\end{eqnarray*}
This equation tells us that the term in the bare action $\int\bar{\theta}\partial_{t}\theta$
does not get renormalized. Moreover the action and the effective action
also enjoy Galilei invariance as we will see in a moment. Thus we
can infer that the term $\int\bar{\theta}\left(\partial_{t}+v^{i}\partial_{i}\right)\theta$
is not renormalized.

We also consider a time dependent shift in the response field $\bar{\theta}\rightarrow\bar{\theta}+\varepsilon\left(t\right)$
and we have 
\begin{eqnarray*}
0 & = & \langle-\int\varepsilon\left(t\right)\left(\partial_{t}\theta\right)+\varepsilon\left(t\right)\langle ff\rangle\bar{\theta}+J_{\bar{\theta}}\varepsilon_{t}\rangle\\
-\frac{\delta\Gamma}{\delta\bar{\theta}} \cdot \varepsilon_{t} & = & -\int\varepsilon\left(t\right)\left(\partial_{t}\theta\right)+\varepsilon\left(t\right)\langle ff\rangle\bar{\theta}\,.
\end{eqnarray*}
This entails the non-renormalization of the term $\frac{1}{2}\bar{\theta}\langle ff\rangle\bar{\theta}$. 

Now we consider the Galilei invariance of the model, the transformation
is:
\begin{eqnarray*}
v\left(t,x\right) & \rightarrow & v^{\prime}\left(t,x^{\prime}\right)+c=v^{\prime}\left(t,x-ct\right)+c\\
x & \rightarrow & x^{\prime}=x-ct\\
t & \rightarrow & t^{\prime}\\
\bar{\theta}\left(t,x\right) & \rightarrow & \bar{\theta}\left(t,x^{\prime}\right)\\
\theta\left(t,x\right) & \rightarrow & \theta\left(t,x^{\prime}\right)\,.
\end{eqnarray*}
This invariance tells us that time derivative actually appears via
Galilei covariant derivative of the form $\partial_{t}+v^{i}\partial_{i}$.
Note that the term quadratic in the velocities is not straightforwardly
invariant. This is due to the fact that the average velocity is set
to zero, in order to see directly Galilei invariance we must express
the kinetic term as $\left(v-\langle v\rangle\right)D_{v}\left(v-\langle v\rangle\right)$.
More precisely it turns out very useful also to consider the following time gauged
Galilei transformation:
\begin{eqnarray*}
v\left(t,x\right) & \rightarrow & v^{\prime}\left(t,x^{\prime}\right)+c\left(t\right)=v^{\prime}\left(t,x-c\left(t\right)\right)+\dot{c}\left(t\right)\\
x & \rightarrow & x^{\prime}=x-c\left(t\right)\\
t & \rightarrow & t^{\prime}\\
\bar{\theta}\left(t,x\right) & \rightarrow & \bar{\theta}\left(t,x^{\prime}\right)\\
\theta\left(t,x\right) & \rightarrow & \theta\left(t,x^{\prime}\right).
\end{eqnarray*}
We observe that
\begin{eqnarray*}
\left(\partial_{t}+v\partial_{x}\right)\theta & \rightarrow & \left[\partial_{t}\theta_{t,x-c\left(t\right)}+\partial_{x}\theta_{t,x-c\left(t\right)}\cdot\left(-\partial_{t}c\right)\right]\\
 &  & +\left(v_{t,x-c\left(t\right)}+\partial_{t}c\left(t\right)\right)\partial_{x}\theta_{t,x-c\left(t\right)}\\
 & = & \left[\left(\partial_{t}+v\partial_{x}\right)\theta\right]^{\prime}.
\end{eqnarray*}
Note that this is a symmetry of the equation (\ref{eq:EOM_theta_force})
itself which is not preserved by the averaging over the velocities.
Indeed performing a transformation in the path integral we obtain
\begin{eqnarray*}
0 & = & \langle-\int vD_{v}\delta v+J_{v}\delta v\rangle\,.
\end{eqnarray*}
This entails the non-renormalization of the ``kinetic'' term for
the velocity fields. Thus we observe that the effective action is invariant under the time--gauged version
of the Galilei transformations except for the quadratic term in the velocity field.
All these non-renormalization theorems are known
in the perturbative framework, see \cite{Adzhemyan1998_incompressible}. 

So far we have been considering symmetries which do not rely on the
presence of the ghost. Indeed $S_{\theta}$ and $S_{gh}$ are both
simultaneously invariant. Now we examine symmetries which exploit
a non-trivial cancellation between $S_{\theta}$ and $S_{gh}$. Let
us note that the action can be rewritten as 
\begin{eqnarray*}
S & = & -\frac{1}{2}\bar{\theta}D_{\bar{\theta}}\bar{\theta}+\bar{\theta}O_{v}\theta+\bar{c}O_{v}c+\frac{1}{2}vD_{v}^{-1}v
\end{eqnarray*}
where $O_{v}\equiv\partial_{t}+v^{i}\partial_{i}-\frac{\kappa}{2}\partial^{2}$.
It is easy to check that $\theta\rightarrow\theta+\varepsilon c,\:\bar{c}\rightarrow\bar{c}-\varepsilon\bar{\theta}$
is a symmetry which we call BRS symmetry in analogy to nomenclature
in \cite{Gozzi:1989bf}. This symmetry has been discussed in \cite{Munoz:1989uk}
and we briefly review it here. The symmetry transformation is linear
and thus the BRS invariance is present also in the effective action.
To derive some useful results we can consider $\langle\delta_{BRS}{\cal O}\rangle=0$
for some particular operators. For instance we can consider $\langle\delta\left(\bar{c}\theta\right)\rangle=\langle\left(-\bar{\theta}\theta+\bar{c}c\right)\rangle=0$
and $\langle\delta\left(\bar{c}\bar{\theta}\right)\rangle\sim\langle\bar{\theta}\bar{\theta}\rangle=0$.
Finally we have also the following symmetry: $\bar{c}\rightarrow\bar{c}-\varepsilon\bar{c}$
and $c\rightarrow c+\varepsilon c$. This symmetry entails the ghost
number conservation.

\section{some frequency integrals \label{sec:Appendix:-some-frequency}}

In this appendix we provide a list of the typical frequency integral
we have encountered. These integral come from the multiplication of
various $\theta\bar{\theta}-$propagator entering in each diagram.
The regularized propagator has the following form:
\begin{eqnarray*}
\frac{1}{\pm i\omega+V} & = & \frac{\mp i\omega+V}{\omega^{2}+V^{2}}\,.
\end{eqnarray*}
First of all we consider
\begin{eqnarray*}
G\left(q\right) & \equiv & \frac{1}{2\pi}\int d\omega\left(-i\omega+\nu q^{2}\right)^{-1}\,.
\end{eqnarray*}
This integral is divergent but its principal value is well defined:

\begin{eqnarray*}
\frac{1}{2\pi}{\cal P}\left[\int d\omega\left(-i\omega+V\right)^{-1}\right] & = & \frac{1}{2\pi}{\cal P}\left[\int d\omega\frac{i\omega+V}{\omega^{2}+V^{2}}\right]\\
 & = & \frac{1}{2}\,.
\end{eqnarray*}
Moreover we need
\begin{eqnarray*}
\frac{1}{2\pi}\int\left(\frac{1}{\pm i\omega+V}\right)^{2} & = & 0\\
\frac{1}{2\pi}\int d\omega\frac{1}{i\omega+V}\frac{1}{-i\omega+V}\frac{1}{i\omega+V} & = & \frac{1}{4}\frac{1}{V^{2}}\,.
\end{eqnarray*}
The first integral above is somewhat unusual in the this type of computations,
since one typically finds
\begin{eqnarray*}
\frac{1}{2\pi}\int\frac{1}{-i\omega+V}\cdot\frac{1}{i\omega+V} & = & \frac{1}{2V}\,.
\end{eqnarray*}
This happens because our diagrams contains not only $\theta\bar{\theta}-$propagator
but also velocity propagators (which are delta in time).


\end{document}